\documentclass[11pt]{article}
\pdfoutput=1

\usepackage{amsmath}

\usepackage{graphics}
\usepackage{graphicx}

\newcommand{\ben}{\begin{enumerate}} 
\newcommand{\een}{\end{enumerate}}

\newcommand{\bef}{\begin{figure}} \newcommand{\eef}{\end{figure}}
\newcommand{\beq}{\begin{equation}} \newcommand{\eeq}{\end{equation}}
\newcommand{\beqr}{\begin{eqnarray}}
\newcommand{\eeqr}{\end{eqnarray}}
\newcommand{\beqrn}{\begin{eqnarray*}}
\newcommand{\eeqrn}{\end{eqnarray*}}
\newcommand{\beqn}{\begin{equation*}}
\newcommand{\eeqn}{\end{equation*}} \newcommand{\bei}{\begin{itemize}}
\newcommand{\eei}{\end{itemize}} \newcommand{\bes}{\begin{small}}
\newcommand{\ees}{\end{small}} \newcommand{\bec}{\begin{center}}
\newcommand{\eec}{\end{center}} 
\newcommand{\tht}{\theta} 
\newcommand{\al}{\alpha} 
\newcommand{\om}{\omega} 
 
\newcommand{\sig}{\sigma} 
\newcommand{\sigt}{\tilde \sigma}

\newcommand{\rt}{\rho_T}

\title{Timescales of spike-train correlation for neural oscillators with common drive}

\author{Andrea K. Barreiro, Eric Shea-Brown, Evan L. Thilo\thanks{Department of Applied Mathematics, University of Washington, 
Box 352420, Seattle, WA 98195}}

\begin{document}
\maketitle

\begin{abstract}  We examine the effect of the phase-resetting curve (PRC) on the transfer of correlated input signals into correlated output spikes in a class of neural models receiving noisy, super-threshold stimulation.  
We use linear response theory to approximate the spike correlation coefficient in terms of moments of the associated exit time problem, and contrast the results for Type I vs. Type II models and across the different timescales over which spike correlations can be assessed.  We find that, on long timescales, Type I oscillators transfer correlations much more efficiently than Type II oscillators.  On short timescales this trend reverses, with the relative efficiency switching at a timescale that depends on the mean and standard deviation of input currents.  This switch occurs over timescales that could be exploited by downstream circuits. 
\end{abstract}

\section{Introduction}

Throughout the nervous system, neurons produce spike trains that are correlated from cell-to-cell.  This correlation, or synchrony, has received major interest because of its impact on how neural populations encode information.  For example, correlations can strongly limit the fidelity of a neural code as measured by the signal-to-noise ratio of homogeneous populations~\cite{zohary94,johnson80,britten92,Bai+01}.  However, the presence or stimulus-dependence of correlations can also enhance coding strategies that rely on discriminating among competing populations~\cite{ad99,alp06,romo03}; in general, the effects of correlation on coding are complex and can be surprisingly strong~\cite{romo03,chen06,roelfsema08,oram98,alp06,SeriesLP04,kohnprep,ad99,shamir04,shamir06,sompolinsky01,schneidman06}.  In addition, stimulus-dependent correlations can modulate or directly carry information directly~\cite{deC+96, samonds03,kohn05,rdsbjr07,Gra+89,biederlack07,chacron08,josicSRD09}.  Correlations also play a major role in how signals are transmitted from layer-to-layer in the brain~\cite{ss00,KuhnAR03,TetzlaffRSAD08}.

What is the origin of correlated spiking?  One natural mechanism is the overlap in the inputs to different neurons -- these common inputs can drive common output spikes.  This poses the question:  how does the process of transferring of input correlations to spike train correlations depend on the nonlinear dynamics of individual neurons?  Such {\it correlation transfer} has been modeled in integrate-and-fire type neurons~\cite{Bin+01,Mor+06,rdsbjr07,brown07,sn98,TetzlaffRSAD08} and, very recently, in phase reductions of neural oscillators~\cite{me08,GalanEU08}.  

In particular,~\cite{me08,GalanEU08} contrast the correlated activity evoked in neural oscillators with Type I (i.e., always positive) vs. Type II (i.e., positive and negative) PRCs. When correlations are measured via equilibrium probability distributions of pairs of neuron phases, or via cross-correlation functions of these phases over time, Type II oscillators display relatively higher levels of correlation~\cite{me08,GalanEU08}.  These results for phase correlation imply a similar finding for spike train correlation in the limit of very short timescales (the connection arises because the spike train cross-correlation function at $\tau = 0$ can be related to the probability that phases will be nearly coincident for the two cells~\cite{pmgh05}.)

In this study, we also contrast correlation transfer in Type I and Type II oscillators (as well as in a continuum of intermediate models).  The primary extension that we make is to study spike train correlation over a range of different timescales.  Specifically, we measure the correlation coefficient $\rt$ between the number of spikes produced by a pair of neurons in a time window of length $T$:
\begin{eqnarray}
\rho_T & = & \frac{Cov(n_1,n_2)}{\sqrt{Var(n_1)}\sqrt{Var(n_2)}} \; \; ;
\end{eqnarray}
Here, $n_1$, $n_2$ are the numbers of spikes output by neurons 1 and 2 
respectively in the time window; see Fig.~1 for an illustration.

\begin{figure}
\begin{center}
\includegraphics[width=5.5in]{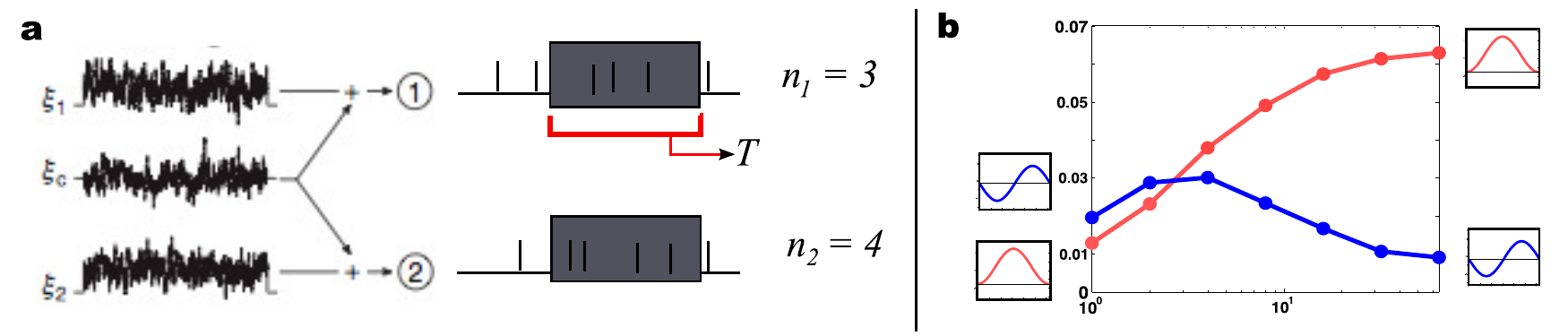}
\end{center}
\caption{(a) A schematic of the setup and correlation metric used in this study. (b) The principal result of our study; that correlation transfer efficiency depends
on both internal dynamics and timescale of readout, with relative efficiency between Type I (light red) and Type II (dark blue) switching as readout timescale changes. 
The graph shows $\rho_T$ for a particular set of model parameters vs. logarithm of the time window $\log(T)$. Insets show
the phase-resetting curves used for the Type I and Type II models.}
\label{fig:setup}
\end{figure}

We first derive a tractable expression for $\rt$ in the long timescale limit $T \rightarrow \infty$ (cf.~\cite{rdsbjr07,brown07}).  This can be given in terms of moments of an associated exit time problem.  This reveals a dramatically higher level of long-timescale correlation transfer in Type I vs. Type II neural oscillator models, the opposite of what was found in the earlier studies over short timescales (see Fig.~1).  Next, we study $\rho_T$ for successively shorter $T$, recovering the earlier findings of \cite{GalanEU08,me08}, and noting the critical timescale below which Type II neurons become more efficient at transferring correlations.  Additional results on how correlation transfer depends on neurons' operating range -- that is, their spike rate and coefficient of variation (CV) of spiking -- are developed as we go along.

\section{Models of neural oscillators and correlation transfer}

\subsection{Phase reductions} 

Neural oscillators can be classified into two types based on their intrinsic dynamics.  Both types have the feature that as applied inputs (or ``injected current") increases, the system transitions from a stable rest state to periodic firing through a bifurcation, the nature of which defines the type~\cite{excit}.  Here, as is often taken to be the case, Type I neurons undergo a saddle-node on invariant circle bifurcation, in which two fixed points collide and disappear, producing a periodic orbit that can have arbitrarily low frequency.  Type II neurons undergo either a subcritical or supercritical Hopf bifurcation, in which a periodic orbit emerges at a non-zero minimum frequency. 

Once the oscillator has passed through this bifurcation, it can be described
by a single equation for its phase, in which inputs are mediated through a
phase-resetting curve (PRC) which indicates the degree to which an input advances
or delays the next spike.  A PRC is derived for a mathematical model by 
phase reduction methods and determined experimentally by repeated perturbations of a system by input ``kicks"~\cite{geomtime}.  Several investigators have demonstrated
a connection between the type of bifurcation and the shape of the PRC:  a Type I oscillator PRC is everywhere positive, so that positive injected current advances the time of the next spike \cite{EI},
whereas a Type II PRC has both positive and negative regions, so positive inputs advance or delay the next spike depending on their timing (~\cite{ermentrout84}; see also~\cite{BHMphase}).  Moreover, the form of Type I and Type II phase-resetting curves \textit{near the bifurcation} 
are given by $(1-\cos \theta)$ and $-\sin(\theta)$ respectively. We investigate these two PRCs, together with a family of PRCs
given by a linear combination of these prototypical examples:
\begin{eqnarray}
Z(\theta) = -\alpha \sin(\theta) + (1-\alpha)(1-\cos(\theta)), \qquad 0 \le \alpha \le 1 \; \;.
\end{eqnarray}
Here $\alpha$ homotopes the PRC from ``purely" Type I to Type II; along the way, intermediate PRCs more representative of phase reductions of biophysical models are encountered ~\cite{EI,BHMphase,HMM93}.

The question of how oscillators with different PRCs synchronize when they are coupled has been the subject of extensive study.  Here, we ask about a different mechanisms by which such oscillators can become correlated.  Specifically, we consider an uncoupled pair of neurons receiving partially correlated noise.
The dynamics have been reduced to a phase oscillator; each neuron is represented by a phase only. 
Each phase $\theta_i$, is governed by the stochastic differential equation
\begin{eqnarray}
d \theta_i = \omega \, dt + \sigma Z(\theta_i) \circ (\sqrt{1-c} \, dW^i_t + \sqrt{c} \,dW^c_t), \qquad \theta_i \in (0,2\pi) \label{eqn:Strat}
\end{eqnarray}
where $\omega,\sigma > 0$, $\circ$ denotes the Stratonovich integral, and
\begin{eqnarray*}
Z(\theta) = -\alpha \sin(\theta) + (1-\alpha)(1-\cos(\theta))\; .
\end{eqnarray*}
Each $\theta_i$ receives independent white noise, $dW^i_t$, and common white noise $dW^c_t$ is received by both. The 
noises are weighted so that the total variance of the noise terms in (\ref{eqn:Strat}) is $\sig^2$.  For the remainder of the paper, we treat the equivalent It\^{o} integral
\begin{eqnarray}
d \theta_i = (\omega + \frac{\sigma^2}{2} Z(\theta_i) Z'(\theta_i) ) \, dt + \sigma Z(\theta_i)  dW_t, \qquad \theta_i \in (0,2\pi). \label{eqn:Ito}
\end{eqnarray}

\subsection{Measuring spike train correlation} 

We record spike times as those times $t_i^k$ at which $\theta_i$ crosses $2\pi$~\cite{EI,BHMphase,ermentrout84}.  Because $Z(2\pi)=0$ and $\om >0$ for all the models we consider, $\theta$ always continues through $2 \pi$ and begins the next period of inter-spike dynamics.   
We consider the output spike trains $y_i(t) = \sum_i \delta(t-t_i^k)$, where
$t_i^k$ is the time of the $k$th spike of the $i$th neuron.  The firing rate of the $i$th cell,
$\langle y_i(t) \rangle$, is denoted $\nu_i$.
As a quantitative measure of correlation over a given time scale $T$, 
we compute the following statistic:
\begin{eqnarray*}
\rho_T & = & \frac{Cov(n_1,n_2)}{\sqrt{Var(n_1)}\sqrt{Var(n_2)}}
\end{eqnarray*}
where $n_1$, $n_2$ are the numbers of spikes output by neurons 1 and 2 
respectively in a time window of length $T$; i.e. $n_i(t) = \int_t^{t+T} y_i(s) \, ds$.

One can show that this is equivalent to
\begin{eqnarray}
\rho_T & = & \frac{\int_{-T}^T C_{12}(t) \frac{T - |t|}{T} \, dt}{\sqrt{\int_{-T}^T C_{11}(t) \frac{T-|t|}{T} \, dt \int_{-T}^T C_{22} (t) \frac{T-|t|}{T} \, dt}} \label{eqn:rho_def_C}
\end{eqnarray}
where $C_{ij} (\tau) = \langle y_i(t) y_j(t + \tau) \rangle - \nu_i \nu_j$~\cite{Cox+66}.  It will be convenient for us to analyze the system in the Fourier domain.
By the Wiener-Khinchin theorem, we can write (\ref{eqn:rho_def_C}) in terms of the power spectra $P_{ij} \equiv \langle \hat{y}^{\ast}_i \hat{y}_j \rangle$ 
as 
\begin{eqnarray}
\rho_T & = & \frac{\int_{-\infty}^{\infty} P_{12}(f) K_T(f)\, df}{\sqrt{\int_{-\infty}^{\infty} P_{11}(f) K_T(f) \, df \int_{-\infty}^{\infty} P_{22} (f) K_T(f) \, df}} \label{eqn:rho_def_P}
\end{eqnarray}
where the kernel $K_T$ is
\begin{eqnarray*}
K_T(f) & = & \frac{4}{T f^2} \sin^2\left( \frac{T f}{2} \right) \; \;.
\end{eqnarray*}

\section{Correlation in the long timescale limit}

\subsection{Linear Response Theory for $\rho_T$}  

We recall the following derivation from \cite{ldl05,rdsbjr07,brown07}. 
Assume that the fraction of noise variance $c$ of the correlated input noise is small;
then we treat the system with common noise as a perturbation to the system
\textit{without} common noise. Because the common noise is small, we will assume that
the response of the system can be treated by linear response theory; that is in
the Fourier domain it can be characterized by
a susceptibility function $A_{\omega,\sigma}(f)$ which gives the scaling factor between
input and response at frequency $f$.

Specifically, we make the \textit{ansatz}~ \cite{ldl05} that the Fourier
transform can be written to lowest order in $c$
\begin{eqnarray*}
\hat{y}_i(f) & = & \hat{y}_{0,i}(f) + \sqrt{c} A_{\omega,\sigma}(f) \hat{Q}(f)
\end{eqnarray*}
where $y_{0,i}$ is the spike output of the neuron without correlated noise,
$\hat{Q}(f)$ is the Fourier transform of the correlated noise, and $A$ is the susceptibility
function. 
Then the cross-spectrum of the two spike trains, $P_{12}(f)$,
satisfies
\begin{eqnarray*}
P_{12}(f) & = & c |A_{\omega,\sigma}(f)|^2 \langle \hat{Q}^{\ast} \hat{Q} \rangle\\
& = & c \sigma^2 |A_{\omega,\sigma}(f)|^2
\end{eqnarray*}
as the base spike trains are independent of each other and the correlated noise $Q$,
and taking the variance of $Q$ to be $\sigma^2$. Then at any finite $T$, $\rho_T$
is linear in $c$, and (cf.~\cite{brown07,Bai+01}):
\begin{eqnarray}
\rho_T(\omega,\sigma) & = & c S_T(\omega,\sigma)\\
& = & c \frac{\sigma^2 \int_{-\infty}^{\infty} |A_{\omega,\sigma}(f)|^2 K_T(f)\, df}{\sqrt{\int_{-\infty}^{\infty} P_{11}(f) K_T(f) \, df \int_{-\infty}^{\infty} P_{22} (f) K_T(f) \, df}} \; \;.
\end{eqnarray}
Note that $S_T(\omega,\sigma)$ multiplies the input correlation $c$ to yield the spike train correlation; for this reason we refer to $S_T(\omega,\sigma)$
as the \underline{correlation gain}.
We recover a simple expression for (\ref{eqn:rho_def_P}) in the limit $T \rightarrow \infty$.
In the numerator, we converge to the value of the integrand at $0$ as $f \rightarrow \infty$:
\[ 
c \sigma^2 |A_{\omega,\sigma} (0)|^2
\]
As $A_{\omega,\sigma}(0)$ is the limit of the susceptibility function as the frequency becomes arbitrarily
small, it must be equivalent to the ratio of the DC response of the system (that is, the firing rate $\nu$) to the strength of a constant DC input. 
Later we will use the symbol $\mu$ to include a DC input explicitly for the purposes of this computation.
Therefore we define
\begin{eqnarray}
\frac{d \nu}{d \mu} & \equiv & A_{\omega,\sigma}(0)
\end{eqnarray}
The denominator converges to $P_{11}(0)$ (assuming the unperturbed oscillators to be 
statistically identical) 
which for renewal processes is simply $CV^2 \nu$.

Putting these results together, as $T \rightarrow \infty$, the finite-time correlations satisfy
\begin{equation}
\lim_{T \rightarrow \infty} \rho_T = c\frac{\sigma^2 (\frac{d\nu}{d\mu})^2}{CV^2 \, \nu} \equiv c S(\omega,\sigma)  \label{eqn:susceptibility}
\end{equation}
where $\nu$ is the mean output firing rate,  $CV$ 
the coefficient of variation
of the interspike intervals, $\sigma$ the input noise amplitude.
Each quantity can be computed from statistics of a single 
oscillator, and combined to yield the long-time correlation gain $S(\omega,\sigma)$.

\subsection{Computing moments of the exit time problem} \label{sec:moments}

The quantities $\nu$, $CV$, and $\frac{d\nu}{d\mu}$ are related by moments of the interspike
intervals (ISI), and as such can be computed using the associated exit time problem. 
Specifically,
\begin{eqnarray}
\nu & =&  \frac{1}{T_1(0)}    \label{eqn:nu} \\ 
CV & = & \frac{\sqrt{(T_2(0) - (T_1(0))^2)}}{T_1(0)} \label{eqn:CV}\\
\frac{d\nu}{d\mu} & = & -\frac{1}{(T_1(0))^2} \frac{d T_1(0)}{d\mu}   \label{eqn:dnu_dmu}
\end{eqnarray}
where $T_1(0)$ is the average time to cross $\theta=2\pi$, given a start at $\theta_0 = 0$ (in other words, 
given a start at the last spike),
and $T_2(0)$ is the second moment of this same quantity.

$T_1(x)$ and $T_2(x)$ are given by solutions of the adjoint equation \cite{gardiner}
\begin{eqnarray}
f_i(x) & =&  A(x) \partial_{x} T_i + \frac{1}{2}B(x) \partial_{x}^2 T_i  \label{eqn:T1}
\end{eqnarray}
where 
\begin{eqnarray}
f_1(x) & = & -1\\
f_2(x) & = & -2 T_1(x)\\
A(x) & = &  \omega + \frac{\sigma^2}{2}Z(x) Z'(x)\\
B(x) & = & \sigma^2 Z(x)^2
\end{eqnarray}
and boundary conditions are given by
$T_i(2 \pi) = 0$, $\frac{\partial T_i}{\partial x}$ bounded at $x = 0$. 
The solution can be obtained by integrating equation~\eqref{eqn:T1} twice over the range  $[0,2\pi]$, using the integrating factor $\Psi$:  
\begin{eqnarray}
\Psi(x) & = & \exp \left( \int^x dx' \frac{2 A(x')}{B(x')} \right), \qquad \Psi(0) = 0 \; \;. \label{eqn:Psi}
\end{eqnarray}
If $\alpha = 0$, then $B(x)$ has no interior zero and we proceed as follows:
\begin{eqnarray}
[\Psi \frac{\partial T_i}{\partial x}]' & = & \frac{2 f_i(x)}{B(x)}\Psi(x)\\
\frac{\partial T_i}{\partial x} & = & \frac{1}{\Psi(x)} \int_0^x \frac{2 f_i(x')}{B(x')} \Psi(x') \, dx' \; \;.  \label{eqn:dTidx}
\end{eqnarray}
If $\alpha > 0$ then $B(x)$ has an interior zero at $\chi(\alpha)$ and we integrate separately on $(0,\chi(\alpha))$ and $(\chi(\alpha), 2\pi)$:
\begin{equation}
\frac{\partial T_i}{\partial x}(x) =
	\begin{cases} 
	\frac{1}{\Psi(x)} \int_0^x \frac{2 f_i(x')}{B(x')} \Psi(x') \, dx' , & x < \chi(\alpha)  \label{eqn:dTidx1}\\
	\frac{1}{\Psi(x)} \int_{\chi(\alpha)}^x \frac{2 f_i(x')}{B(x')} \Psi(x') \, dx' , & x > \chi(\alpha)
	\end{cases}
\end{equation}
Finally $T_i(0)$ is given by a second integration:
\begin{eqnarray}
T_i(0) & = & \int_{2\pi}^0 \frac{\partial T_i}{\partial x}(x) \, dx \; \;. \label{eqn:Ti0}
\end{eqnarray}

Here, the argument of the exponential function in $\Psi(x)$ can be any antiderivative of $\frac{2 A(x)}{B(x)}$.
$\Psi(0)$ (and $\Psi(\chi(\alpha))$, if $\alpha > 0$) is zero because the adjoint equation (\ref{eqn:T1}) has an irregular singularity at
$x = 0$ (and at $x = \chi(\alpha))$; consequently
$\frac{2A(x)}{B(x)} \rightarrow \infty$ and $\int^x dx \frac{2A(x)}{B(x)} \rightarrow -\infty$ as $x \rightarrow 0^{+}$ (and as $x \rightarrow \chi(\alpha)^{+}$).    
This accounts for the difference between (\ref{eqn:Ti0}) and, for example,
Eqn. (5.2.157) in \cite{gardiner} \footnote{The expression we obtain is identical to the evaluation of this expression in the case of 
a reflecting boundary at $0$.}. Note that, also because $Z(0) = 0$ and $\omega > 0$,
$x = 0$ is an entrance boundary, so any exit must take place at $2 \pi$ (i.e.
an exit from the interval $(0,2\pi)$ is equivalent to a spike).

For the class of PRCs that we consider, the antiderivative in (\ref{eqn:Psi}) can be evaluated symbolically, 
therefore $\Psi(x)$ can be evaluated analytically. The integrals in (\ref{eqn:dTidx}, \ref{eqn:Ti0})
must be evaluated by numerical quadrature.
 
We next discuss the integrability of (\ref{eqn:dTidx}) and (\ref{eqn:dTidx1}).  
First, we consider the case of the theta neuron ($\alpha = 0$), for which $B(x)$ has only two zeros; at $0$ and $2\pi$. We can confirm the integrability by checking the following conditions:
\begin{eqnarray}
\lim_{x\rightarrow 0^+} \int_0^x \frac{2f(x')}{B(x')} \Psi(x') dx' & = & 0  \label{eqn:lHop_cond}\\
\lim_{x\rightarrow 0^+} \Psi(x) & = & 0\\ 
\lim_{x\rightarrow 2\pi^-} \int_0^x \frac{2f(x')}{B(x')} \Psi(x') dx' & = & \pm \infty\\
\lim_{x\rightarrow 2\pi^-} \Psi(x) & = & \infty \label{eqn:lHop_cond_last}
\end{eqnarray}
If these are satisfied then by l'Hospital's rule,
\begin{eqnarray}
\frac{\partial T_i}{\partial x} & = & \frac{1}{\Psi(x)} \int_0^x \frac{2f_i(x')}{B(x')} \Psi(x') dx'
\end{eqnarray}
is finite at the endpoints; in fact
\begin{eqnarray}
\lim_{x\rightarrow 0} \frac{\partial T_i}{\partial x} & = & \lim_{x\rightarrow 0} \frac{f_i(x)}{A(x)}\\
\lim_{x\rightarrow 2\pi} \frac{\partial T_i}{\partial x} & = & \lim_{x\rightarrow 2\pi} \frac{f_i(x)}{A(x)}
\end{eqnarray}
and we can use a quadrature method that can handle integrable singularities. 
For $\alpha > 0$, $B(x)$ has one interior zero, dividing the domain into two intervals
on which (\ref{eqn:T1}) has irregular singularities at each end. 
$\Psi$ must be computed separately on each interval. Again we find that $\frac{\partial T_i}{\partial x}$ is finite on each interval, 
permitting computation of (\ref{eqn:dTidx1}) with a standard quadrature routine. 

Finally, we compute the derivative of the firing rate with respect to a DC input; i.e.
$\frac{\partial \nu}{\partial \mu}$ for the system
\begin{equation}
d \theta = \omega \, dt + Z(\theta)(\mu \, dt + \sigma \circ dW_t), \qquad \theta \in [0,2\pi)
\end{equation} 
which is equivalent to the It\^{o} SDE
\begin{equation}
d \theta = (\omega + \frac{\sigma^2}{2} Z(\theta) Z'(\theta) + \mu Z(\theta)) \, dt  + \sigma Z(\theta) dW_t, \qquad \theta \in [0,2\pi)   \; . \label{eqn:sde_with_mu}
\end{equation}
We wish to differentiate $\nu$ with respect to $\mu$ and evaluate at $\mu = 0$. According to equation~(\ref{eqn:dnu_dmu}) we must evaluate
$\frac{d T_1}{d\mu}(0,\mu)$ where the drift term $A(x,\mu) = \omega + \frac{\sigma^2}{2}Z(x) Z'(x) + \mu Z(x)$ is a function of both $x$ and $\mu$.
In general, $T_1$, $T_2$ and $\Psi$ are also functions of two arguments (e.g. $T_1(x,\mu)$) and we have indicated the arguments where needed for clarity.
The notation $\frac{\partial}{\partial x}$ will refer to differentiation with respect to the first argument.

We first consider the case $\alpha = 0$.
Rewriting ({\ref{eqn:Ti0}), we have
\begin{equation}
T_1(0,\mu) = \int_{2\pi}^0 \frac{1}{\Psi(x,\mu)} \int_0^x \frac{-2}{B(x')} \Psi(x',\mu) dx'  \, dx \label{eqn:dT1_start}
\end{equation}
where 
\begin{equation}
\Psi(x,\mu) = \exp \left[ \int^x \frac{2A(x',\mu)}{B(x')} dx' \right] \; .
\end{equation}

\noindent Therefore,
\begin{eqnarray}
\frac{\partial T_1}{\partial \mu}(0,\mu) & = & \frac{d}{d\mu} \left( \int_{2\pi}^0 \int_0^x \frac{-2}{B(x')} \frac{\Psi(x',\mu)}{\Psi(x,\mu)} dx' dx \right)  \nonumber \\
& = & \int_{2\pi}^0 \int_0^x \frac{-2}{B(x')} \frac{\Psi_{\mu}(x',\mu) \Psi(x,\mu) - \Psi_{\mu}(x,\mu) \Psi(x',\mu)}{\Psi(x,\mu)^2} dx' dx \label{eqn:dvdu_zero}
\end{eqnarray}
We use the relationship 
\begin{equation}
\Psi_{\mu}(x,\mu) = \Psi(x,\mu) \int_0^x \frac{2}{B(y)}\frac{\partial A(y,\mu)}{\partial \mu} dy
\end{equation}
to find that 
\begin{eqnarray}
\frac{\partial T_1}{\partial \mu}(0,\mu) & = &  \int_{2\pi}^0 \int_0^x \frac{2}{B(x')} \frac{\Psi(x',\mu)}{\Psi(x,\mu)} \int_{x'}^x \frac{2}{B(y)}\frac{\partial A(y,\mu)}{\partial \mu} dy \, dx' \, dx  \nonumber \\
& = &  \int_{2\pi}^0 \int_0^x \int_0^y \frac{2}{B(x')} \frac{\Psi(x',\mu)}{\Psi(x,\mu)}\frac{2}{B(y)}\frac{\partial A(y,\mu)}{\partial \mu} dx' \, dy \, dx \label{e.derint}
\end{eqnarray}
assuming that the order of integration over $x'$ and $y$ can be switched. By Tonelli's theorem, this is valid if the integrand is
single-signed. For $\alpha = 0$, the integrand is always nonnegative (note that $\frac{\partial A}{\partial \mu}(x) = Z(x)$ and that $B(x)$ is nonnegative). 

Next, we establish the integrability of the expression~\eqref{e.derint}.  Rewriting, we have
\begin{eqnarray*}
\frac{\partial T_1}{\partial \mu}(0,\mu) & = & \int_{2\pi}^0 \frac{1}{\Psi(x,\mu)} \int_0^x \frac{2}{B(y)}\frac{\partial A}{\partial \mu} \Psi(y,\mu) \times \frac{1}{\Psi(y,\mu)} \int_0^y \frac{2}{B(x')}\Psi(x',\mu) dx' \, dy \, dx\\
& = & \int_{2\pi}^0 \frac{1}{\Psi(x,\mu)} \int_0^x \frac{2}{B(y)}\frac{\partial A}{\partial \mu} \Psi(y,\mu) \times \frac{\partial T_1}{\partial x}(y) dy \, dx
\end{eqnarray*}
As $\frac{\partial A}{\partial \mu}(x) = Z(x)$ and $\frac{\partial T_1}{\partial x}$ are bounded, we can use the same conditions for
integrability unchanged from (\ref{eqn:lHop_cond} - \ref{eqn:lHop_cond_last}).

A parallel derivation to (\ref{eqn:dT1_start} - \ref{e.derint}) can be made when $\alpha > 0$. In this case we have
\begin{eqnarray}
\frac{\partial T_1}{\partial \mu}(0,\mu) & = & \int_{2\pi}^0 \frac{\partial}{\partial \mu} \left( \frac{\partial T_1}{\partial x}(x,\mu) \right) \, dx  \label{eqn:T1_alpha1}
\end{eqnarray}
where we have used the boundary condition $T_1(2\pi,\mu) = 0$. The integrand is given by
\begin{equation}
\frac{\partial}{\partial \mu} \frac{\partial T_1}{\partial x} (x,\mu) =
\begin{cases}
	\int_0^x \int_0^y \frac{2}{B(x')} \frac{\Psi(x',\mu)}{\Psi(x,\mu)}\frac{2}{B(y)}\frac{\partial A(y,\mu)}{\partial \mu} dx' \, dy, & x < \chi(\alpha) \label{eqn:T1_alpha2a}\\
        \int_{\chi(\alpha)}^x \int_{\chi(\alpha)}^y \frac{2}{B(x')} \frac{\Psi(x',\mu)}{\Psi(x,\mu)}\frac{2}{B(y)}\frac{\partial A(y,\mu)}{\partial \mu} dx' \, dy, & x > \chi(\alpha)
        \end{cases}
\end{equation}
Here, the switch in the order of integration is justified by the fact that the
integrand $\frac{\partial A}{\partial \mu}$ is positive on $(0,\chi(\alpha))$ and negative on $(\chi(\alpha),2\pi)$, while the range of
integration in Eqn. (\ref{eqn:T1_alpha2a}) is always restricted to lie in one region or the other.

\subsection{Patterns of correlation transfer over long timescales}\label{s.rholong}

\begin{figure}[ht]
\begin{center}
\includegraphics[width=5.5in]{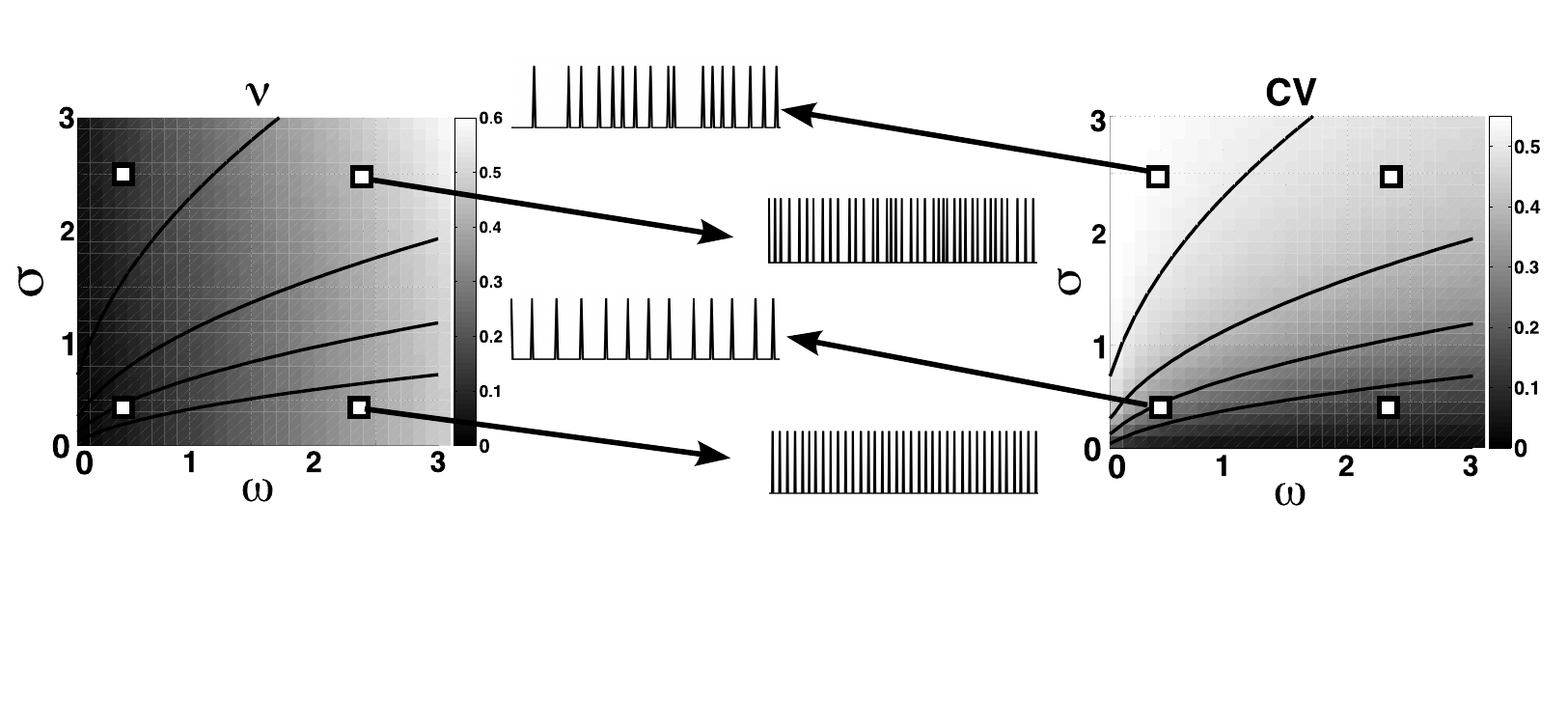}
\end{center}
\caption{Firing rate (left) and CV (right) of the theta model neuron over a range of parameters
$\omega$ and $\sigma$. Spike trains are illustrated for four sets of $(\omega,\sigma)$ values (see white squares),
notably mean-driven (high $\omega$, low $\sigma$ - bottom spike train) and fluctuation-driven 
(low $\omega$, high $\sigma$ - top spike train). Level sets of $\tilde{\sigma}$ are plotted (see text).}
\label{fig:nu_and_CV}
\end{figure}

\begin{figure}
\begin{center}
\includegraphics[width=5.5in]{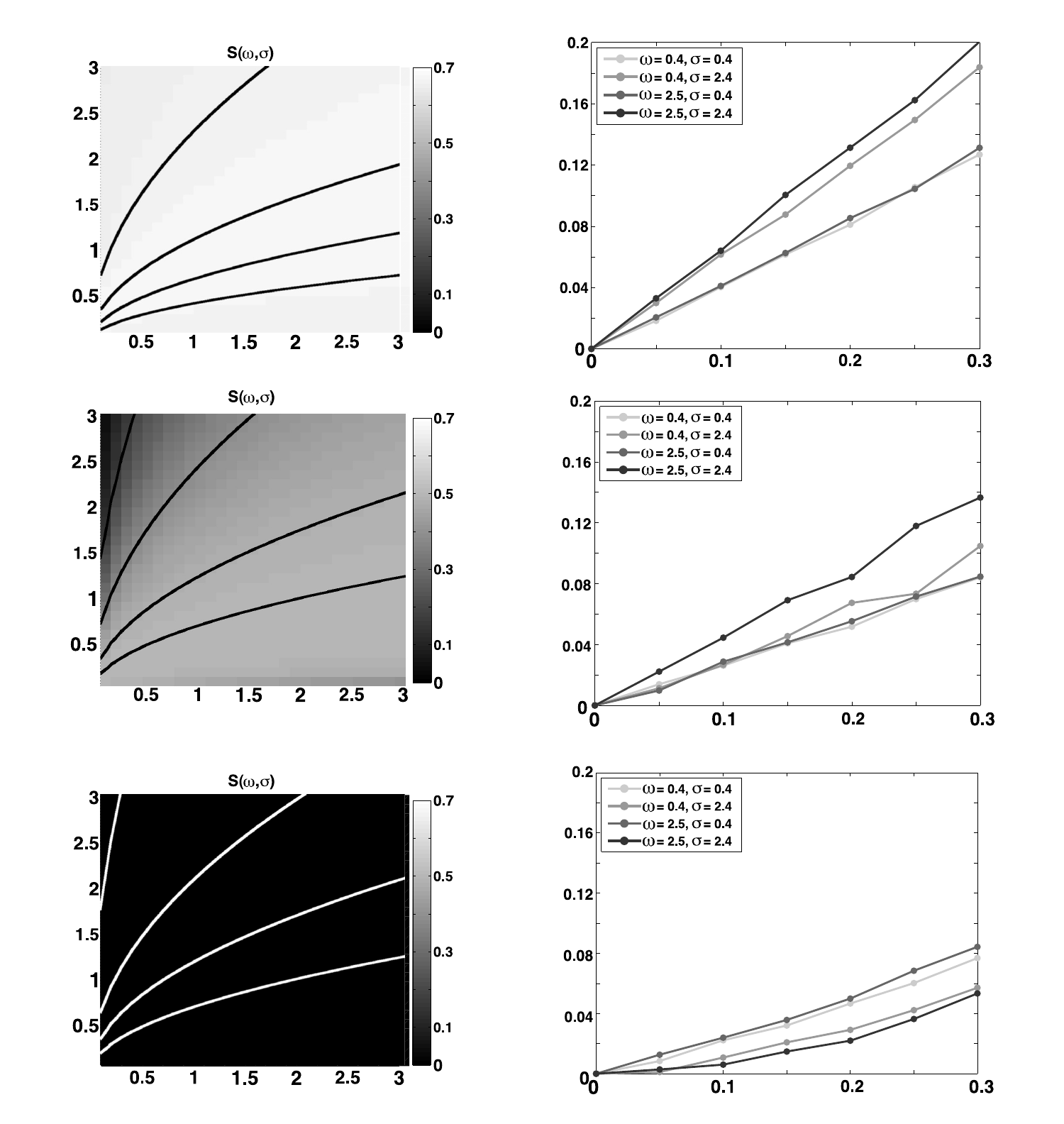}
\end{center}
\caption{(Left) Susceptibility for $\alpha = 0$ (top), $\alpha = 0.5$ (middle), and $\alpha = 1$ (bottom) over a range of parameters $\omega$ and $\sigma$;
lines are level sets of $\tilde{\sigma} = \sigma/\sqrt{\omega}$. 
(Right) $\rho_T$ vs. $c$ from Monte Carlo simulations for $\alpha = 0$ (top), $\alpha = 0.5$, and $\alpha = 1$ (bottom). Specific $(\omega, \sigma)$
values shown as in Figure \ref{fig:nu_and_CV}.}
\label{fig:rho_inflimit}
\end{figure}

Having derived and shown how to evaluate formula~\eqref{eqn:susceptibility}, we next use it to evaluate spike count correlations $\rho$.  The results are formally valid in the limits of timescale $T \rightarrow \infty$ and input correlation $c \rightarrow 0$, so we also conduct Monte Carlo simulations to reveal behavior of $\rho_T$ for large but finite $T$ and intermediate input correlation up to $c=0.3$, and to test the applicability of our formula in these regimes.

We compute these spike correlations over a range of $\omega$ and $\sigma$ values that explores a full dynamical regime of the model.  By this we mean that the values we use span from dominantly mean-driven firing ( e.g. $\om = 2.5$, $\sig = 0.4$ at lower-right white square in Fig. ~\ref{fig:nu_and_CV}), to dominantly fluctuation-driven firing (e.g. $\om=0.4$, $\sig =2.4$ at upper-left white square), and all intermediate possibilities. 

The resulting output firing rate $\nu$, computed via Eqn.~\eqref{eqn:nu},  ranges from $0$ to $0.9$ (measured in spikes per time unit); see Fig.~\ref{fig:nu_and_CV} (left panel).  Note that $\nu$  increases strongly with $\omega$ and only weakly with $\sigma$ (see Sec.~\ref{s.ana}).  The CV (via Eqn.~\eqref{eqn:CV}) ranges from $0$ to $0.55$ (Fig.~\ref{fig:nu_and_CV}, right panel).  We were unable to reach higher values of CV even with a much expanded range of $\sigma$.  By using a time change in the equations, CV can be seen to depend on the input parameters $\omega$ and $\sigma$
only through the relationship $\tilde{\sigma}=\sigma/\sqrt{\omega}$ (see later in this section)  and is therefore invariant on level curves of this ratio. As Fig.~\ref{fig:nu_and_CV} shows, CV increases with this ratio. 

We first fix a moderately long time window $T=32$, corresponding to $1.6-16$ interspike intervals for the parameter range at hand, and compute $\rho_T$ from Monte Carlo simulations for a range of correlation strengths $c \in [0, 0.3]$; see Fig.~\ref{fig:rho_inflimit} (right column).  We see that, for Type I models, $\rho_T$ is close to linear in this range of $c$, as for the linear response theory.  For Type II models, linearity holds over a decreased range of $c$.  Additionally, note that the limiting formula~Eqn. (\ref{eqn:susceptibility}) for $\rho$ gives a close approximation to $\rho_{T=32}$ for Type I models in more fluctuation-driven regimes.  The approximation is worse for dominantly mean-driven firing, and for Type II models, but the trend that correlations are lower for Type II models is correctly predicted by Eqn.~\ref{eqn:susceptibility}. 

Next, we discuss the trends in $\rho$ predicted by the linear response theory.  Two findings stand out in the left hand panels of Fig.~\ref{fig:rho_inflimit}.  First, values of 
$S(\omega,\sigma)$ (and hence $\rho \approx S \,c$) are much larger for Type I ($\alpha = 0$) than for Type II ($\alpha=1$) models.  Second, $S(\omega,\sigma)$ is nearly constant as the input mean and standard deviation $\omega$ and $\sigma$ vary over a wide range, for both the Type I and Type II models.  In sum, Type I models transfer $\approx 66 \%$ of their input correlations into spike correlations over long timescales; Type II models transfer {\it none} of these input correlations, producing long-timescale spike counts that are uncorrelated.  Additionally, an intermediate model ($\alpha=1/2$) transfers an intermediate level of correlations, and these levels {\it do} depend on $\omega$ and $\sigma$.  We will provide a partial explanation for overall trends in correlation transfer with $\alpha$ in Section~\ref{s.ana}.

\begin{figure}[ht]
\begin{center}
\includegraphics[width=5.0in]{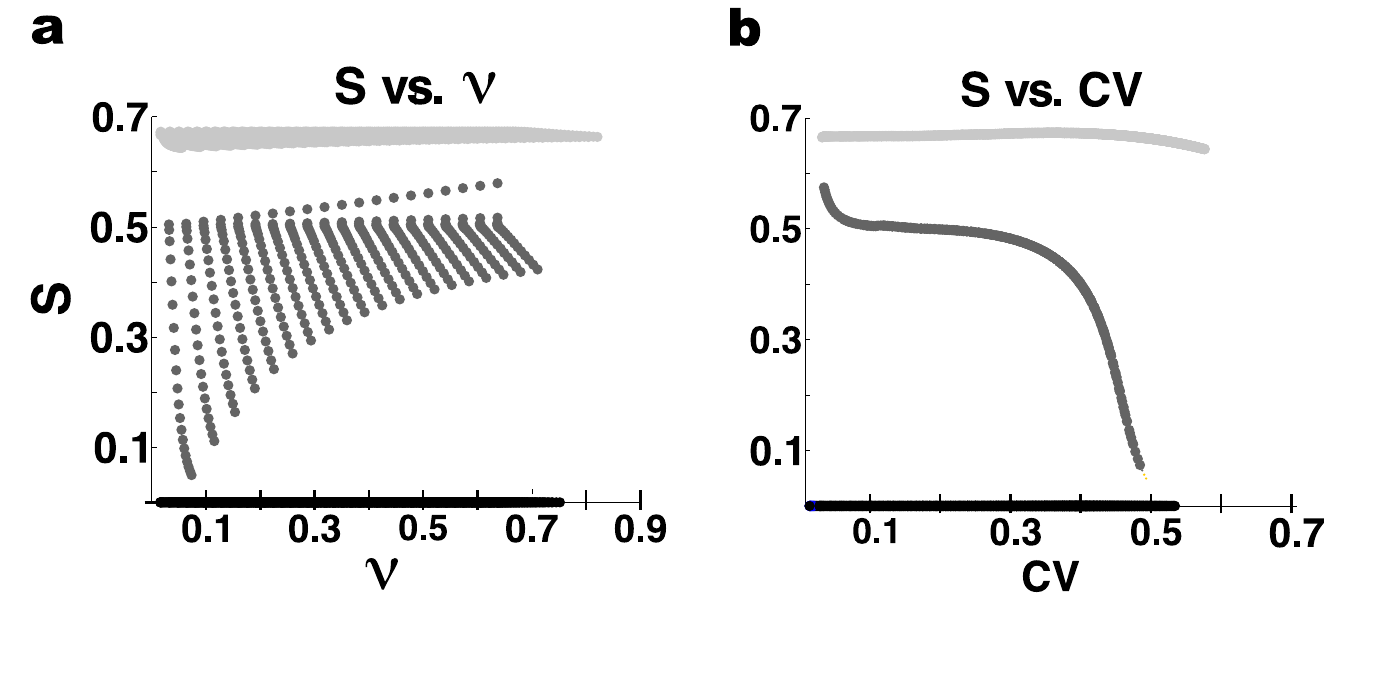}
\end{center}
\caption{Susceptibility vs. $\nu$ (a) and susceptibility vs. $CV$ (b) for $\alpha = 0$ (light gray), $0.5$ (medium gray), $1$ (black).}
\label{fig:CV_dependence}
\end{figure}

Fig.~\ref{fig:CV_dependence} provides an alternative view of these results, by plotting the correlation gain $S$ vs. the firing rate and CV that are evoked by input parameters drawn from the whole range of $\omega,\; \sigma$.  First, note that $S$ does not vary with firing rate for the Type I or Type II models, as expected from the previous plots.  For the intermediate ($\alpha = 1/2$) model, $S$ does not display a clear functional relationship with firing rate, but there is such a relationship with CV (Fig.~\ref{fig:CV_dependence}(b)).  We note that all of these findings for the phase oscillators under study are in contrast to the behavior of linear integrate and fire neurons, which produce a strongly increasing, nearly functional relationship with firing rate~\cite{rdsbjr07,brown07}; we revisit this point in the discussion. 

%

Now, we discuss a scaling relationship for the underlying equations that simplifies the parameter dependence and helps to explain the plots of $S$ vs. $\nu$ and $S$ vs. CV.  This symmetry (which was noted in \cite{lindner03} for the quadratic integrate and fire model), allows us to reduce the
free parameters $\omega, \sigma$ to one parameter $\tilde{\sigma} \equiv \sigma/\sqrt{\omega}$.
The stochastic differential equation
\begin{eqnarray*}
d\theta_t & = & \left( \omega + \frac{\sigma^2}{2} Z(\theta)Z'(\theta) + \mu Z(\theta) \right) dt + \sigma Z(\theta) dW_t
\end{eqnarray*}
becomes, under a time change $\tau = \omega t$,
\begin{eqnarray}
d\theta_{\tau} & = &  \left( 1 + \frac{\sigma^2}{2\omega} Z(\theta)Z'(\theta) + \frac{\mu}{\omega} Z(\theta) \right) d\tau + \frac{\sigma}{\sqrt{\omega}} Z(\theta) dW_{\tau}\\
& = & \left( 1 + \frac{\tilde{\sigma}^2}{2} Z(\theta)Z'(\theta) + \frac{\mu}{\omega} Z(\theta) \right) d{\tau} + \tilde{\sigma} Z(\theta) dW_{\tau} \label{eqn:tilde_sigma}
\end{eqnarray}
Each exit time must scale identically under this transformation, so that the exit time moments scale as 
\begin{eqnarray}
\tilde{T}_1 & = & \omega T_1 \rightarrow \tilde{\nu} = \frac{\nu}{\omega}\\
\tilde{T}_2 & = & \omega^2 T_2
\end{eqnarray}
and therefore the CV $=\sqrt{\tilde{T}_2 - \tilde{T}_1^2}/\tilde{T}_1$ is invariant under the time change.  Thus, the CV, which is computed with $\mu=0$, depends only on $\sigt$, not on $\om$ and $\sig$ separately.  Now consider the two sets of parameters $(\omega, \sigma)$ and $(1, \tilde{\sigma})$, such that
$\tilde{\sigma} = \sigma/\sqrt{\omega}$.
According to (\ref{eqn:tilde_sigma}), the effect of $\mu$ is scaled by $1/\omega$ so that 
\begin{eqnarray}
\frac{d\tilde{\nu}}{d\mu} & = & \frac{1}{\omega}\frac{d\nu}{d\mu} \; .
\end{eqnarray}
Therefore, the susceptibility
\begin{eqnarray}
S(\omega,\sigma) & = & \frac{\sigma^2 (\frac{d\nu}{d\mu})^2}{CV^2 \, \nu}\\
& = & \frac{\sigma^2 \frac{1}{\omega^2}(\frac{d\tilde{\nu}}{d\mu})^2}{\tilde{CV}^2 \, \frac{\tilde{\nu}}{\omega}}\\
& = & \frac{\sigma^2}{\omega} \frac{(\frac{d\tilde{\nu}}{d\mu})^2}{\tilde{CV}^2 \, \tilde{\nu}}\\
& = & S(1,\tilde{\sigma})
\end{eqnarray}
is invariant under the change of parameters $(\omega,\sigma) \rightarrow (1, \sigma/\sqrt{\omega})$;
exactly the same change of parameters under which the $CV$ is conserved.
Therefore, if there is a single contour for each value of $CV$ (i.e., there are no disconnected contours), we expect that $S$ will be a function of CV, as in Fig.~\ref{fig:CV_dependence}(b).  Conversely, note that the firing rate will vary widely over any particular $\tilde{\sigma}$ contour, so that many different firing rates can be expected to yield the same $S$; unless $S$ is constant, we expect a given firing rate to be associated with a range of $S$ values, as also seen in Fig.~\ref{fig:CV_dependence} (a).

\begin{figure}[ht]
\begin{center}
\includegraphics[height=2in]{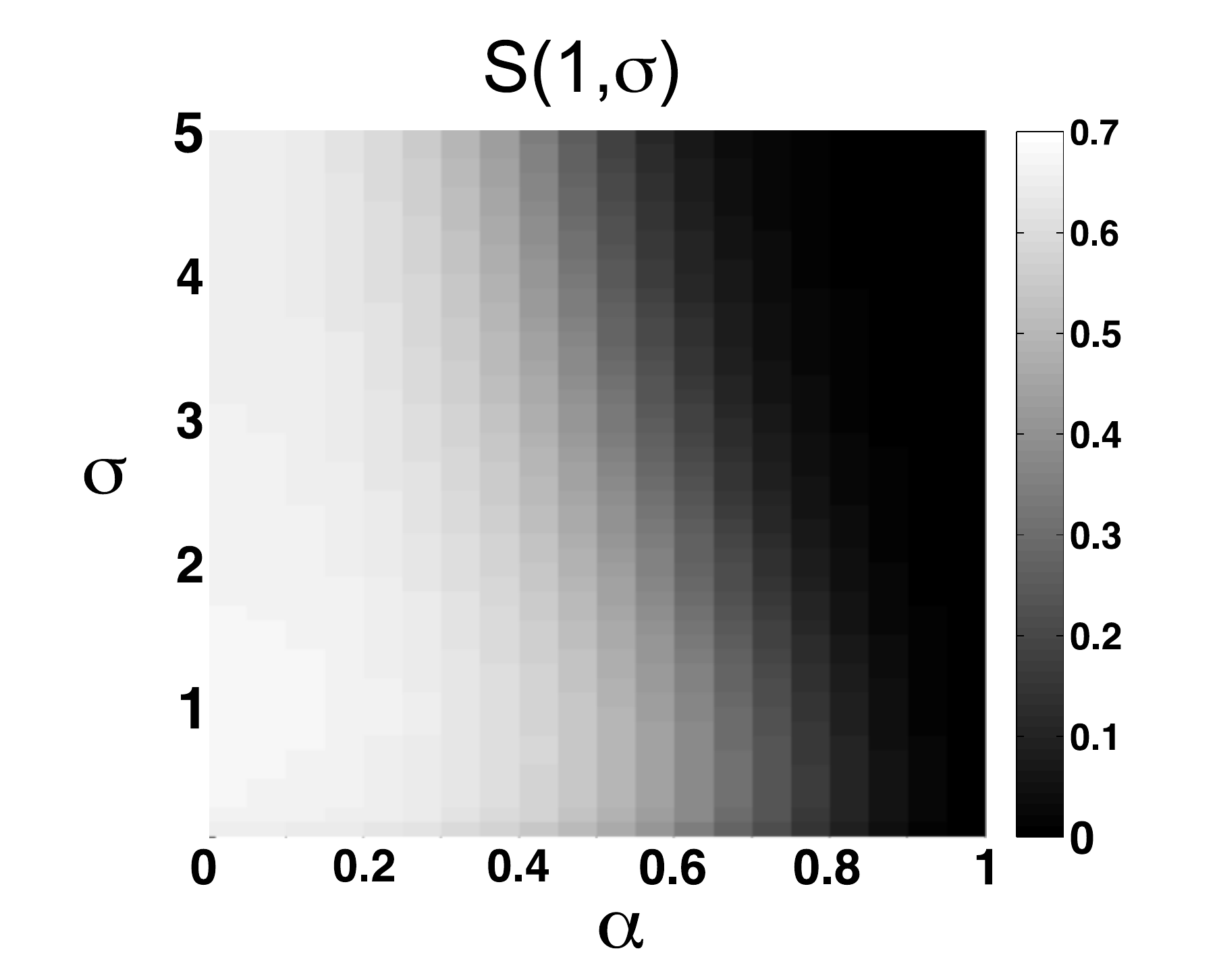}
\end{center}
\caption{$S(1,\sigma)$ for a continuum of models $\alpha \in [0,1]$. }
\label{fig:alpha_switch}
\end{figure}

Finally, in Fig. \ref{fig:alpha_switch}, we demonstrate that the behavior of $S$ seems to roughly ``interpolate" between the $\alpha=0$, $1/2$, 1 cases considered here as it varies over it whole range of $\alpha$. $S$ typically decreases as the total noise variance $\sigma$ increases.

\subsection{Analytical arguments for Type I vs. Type II difference in correlation gain $S$}~\label{s.ana}


We now seek to explain the drop in $S(\omega,\sigma)$ (Eqn.~(\ref{eqn:susceptibility})) as $\alpha$ increases, ranging from Type I ($\al=0$) to Type II ($\al=1$) PRCs.   There are two tractable limits in which explicit calculations can be performed.  First, we show that $S(\omega,\sigma)=0$ for ``purely" Type II models, for any values of $\omega$ and $\sigma$.  Next, for arbitrary $\alpha$, we derive an $S(\omega,\sigma)$ valid to first order in $\sigma^2$ -- this reveals a monotonic decrease in $S$ with $\alpha$ and gives a good description of correlation transfer even for $\sigma \approx 1$.  Finally, we given an intuitive argument to buttress these calculations.

\subsubsection*{$S=0$ for purely Type II models}

We start with $\alpha=1$.  It is straightforward to see that $d\nu/d\mu = 0$ for any $\om,\,\sigma$. Recall that $d\nu/d\mu$ is given by the integral
of a function over the interval $(0,2\pi)$, in Eqn.~(\ref{eqn:T1_alpha1}); we will show this function integrates to zero.
Rewriting Eqn.~(\ref{eqn:T1_alpha2a}),
using $\chi(1) = \pi$,
\begin{equation}
\frac{\partial}{\partial \mu} \frac{\partial T_1}{\partial x}  = 
	\begin{cases}
 	\frac{1}{\Psi(x,\mu)} \int_0^x \frac{2}{B(y)}\frac{\partial A}{\partial \mu}(y,\mu) \Psi(y,\mu) \times \frac{\partial T_1}{\partial x}(y,\mu) \, dy, & x < \pi\\
 	\frac{1}{\Psi(x,\mu)} \int_{\pi}^x \frac{2}{B(y)}\frac{\partial A}{\partial \mu}(y,\mu) \Psi(y,\mu) \times \frac{\partial T_1}{\partial x}(y,\mu) \, dy, & x > \pi.
	\end{cases}
\end{equation}
$\Psi(x,\mu)$,$B(x)$, and $\frac{\partial T_1}{\partial x}(x,\mu)$ are each $\pi$-periodic (i.e. $f(x) = f(x+\pi)$). $\frac{\partial A}{\partial \mu} \equiv Z(x) = -\sin(x)$, however, is
anti-periodic ($f(x) = -f(x+\pi)$). Then 
\begin{eqnarray}
\frac{\partial}{\partial \mu} \frac{\partial T_1}{\partial x}(x,\mu) & = & - \frac{\partial}{\partial \mu} \frac{\partial T_1}{\partial x}(x+\pi,\mu)
\end{eqnarray}
and integrating $\frac{\partial}{\partial \mu} \left( \frac{\partial T_1}{\partial x}(x,\mu) \right)$ over a full period $x \in (0,2\pi)$ yields zero.  Plugging this into Eqn.~(\ref{eqn:susceptibility}), we see that  $S(\om,\sig) \equiv 0$ for models with Type II PRCs $Z(\tht)=-\sin(\tht)$ ($\nu$ and $CV$ are always nonzero in this paper).

\subsection*{Evaluation of $S(\omega,\sigma)$ in the limit $\sigma \rightarrow 0$}
 We next derive an analytical expression for $S(\omega,\sigma)$ in the limit $\sigma \rightarrow 0$. 
 We will show that each relevant term
$T_1(0)$, $T_2(0)$, and $\frac{\partial T_1}{\partial \mu}(0)$ admits an asymptotic expansion in the small
parameter $\sigma^2$. We compute these terms explicitly and combine them to get the first term of the associated
expansion for $S(\omega,\sigma)$.

 If we examine the integral in (\ref{eqn:dTidx}) and the inner integral of (\ref{eqn:dvdu_zero}) for $\alpha = 0$, 
 or the integrals in (\ref{eqn:dTidx1}), and (\ref{eqn:T1_alpha2a}) for $\alpha > 0$, 
 we see that we can write each of them in the form 
 \begin{eqnarray}
 \frac{1}{Z(y)} \int_a^y f(x') \exp \left( \frac{1}{\sigma^2}(F(x')-F(y)) \right) dx' \label{eqn:asymp_Ti}
 \end{eqnarray}
where $F(x')$ is strictly increasing on $(a,y)$. $a$ may be either $0$ or $\chi_{\alpha}$, depending on the circumstance.
This integral admits an asymptotic expansion in $\sigma^2$, with successive terms essentially
given by integrating by parts and retaining only the contribution from the end of the interval where $F(x')-F(y) = 0$.
In the case of $T_1(x)$ we have 
\begin{eqnarray*}
f(y) & = & \frac{1}{Z(y)}.
\end{eqnarray*}
In the case of $T_2(x)$ we have 
\begin{eqnarray*}
f(y) & = & \frac{T_1(y)}{Z(y)}
\end{eqnarray*}
and for $d\nu/d\mu$ 
\begin{eqnarray*}
f(y) & = &  \frac{\partial T_1}{\partial x}(y).
\end{eqnarray*}
In each case $F(x')$ is given by the antiderivative of $\frac{2\omega}{Z(x')^2}$; because this
is a positive function on $(0,\chi(\alpha))$ and $(\chi(\alpha),2\pi)$, $F(x')$ must clearly be increasing
on these intervals. 
The integral
\begin{eqnarray}
\int_a^b f(x') \exp \left( \lambda \phi(x') \right) dx'  \label{eqn:asymp_phi}
\end{eqnarray}
admits the following expansion for $\lambda \rightarrow \infty$, if $\phi' > 0$ on $[a,b]$ and $f$ meets certain conditions 
(see, for example, \cite{bleistein}):
\begin{eqnarray}
I(\lambda) & \sim & \sum_{n=0}^{\infty} \frac{(-1)^n}{\lambda^{n+1}} \left[ \exp \left( \lambda \phi(x) \right) \left[ \frac{1}{\phi'(x)} \frac{d}{dx} \right]^n \frac{f(x)}{\phi'(x)} \right]^b_a
\end{eqnarray}
If $\phi(x) \rightarrow -\infty$ as $x \rightarrow a$, as is the case in our use of the formula,
only the right-hand endpoint makes a contribution to the integral.
 
Substituting this contribution into the outer integral and evaluating these quantities to the required order, we find
\begin{eqnarray}
T_1(0) & = & \frac{2\pi}{\omega} + O(\sigma^4)\\
T_2(0) & = & \frac{4\pi^2}{\omega^2} + \sigma^2 \left(\frac{\pi}{\omega^3} \left( 3 - 6\alpha + 4 \alpha^2 \right) \right) + O(\sigma^4)\\
\frac{dT_1}{d\mu}(0) & = & -\frac{2\pi}{\omega} \left( 1-\alpha \right) + O(\sigma^4)
\end{eqnarray}
In passing, we note that the firing rate gain, $d\nu/d\mu$, is given by
\begin{eqnarray}
\frac{d\nu}{d\mu} & = & -\frac{1}{T_1(0)^2}\frac{dT_1}{d\mu}(0) \\
& = & \frac{1-\alpha}{2\pi} + O(\sigma^4)
\end{eqnarray}
Putting these results together we see that
\begin{eqnarray}
S(\omega,\sigma) & = & \frac{2 (1-\alpha)^2}{3-6\alpha + 4\alpha^2} + O(\sigma^2)
\end{eqnarray}
It can be readily checked that this function decreases monotonically from a value of
$2/3$ at $\alpha = 0$, to a value of $0$ at $\alpha = 1$.
\begin{figure}[h]
\begin{center}
\includegraphics[height=2in]{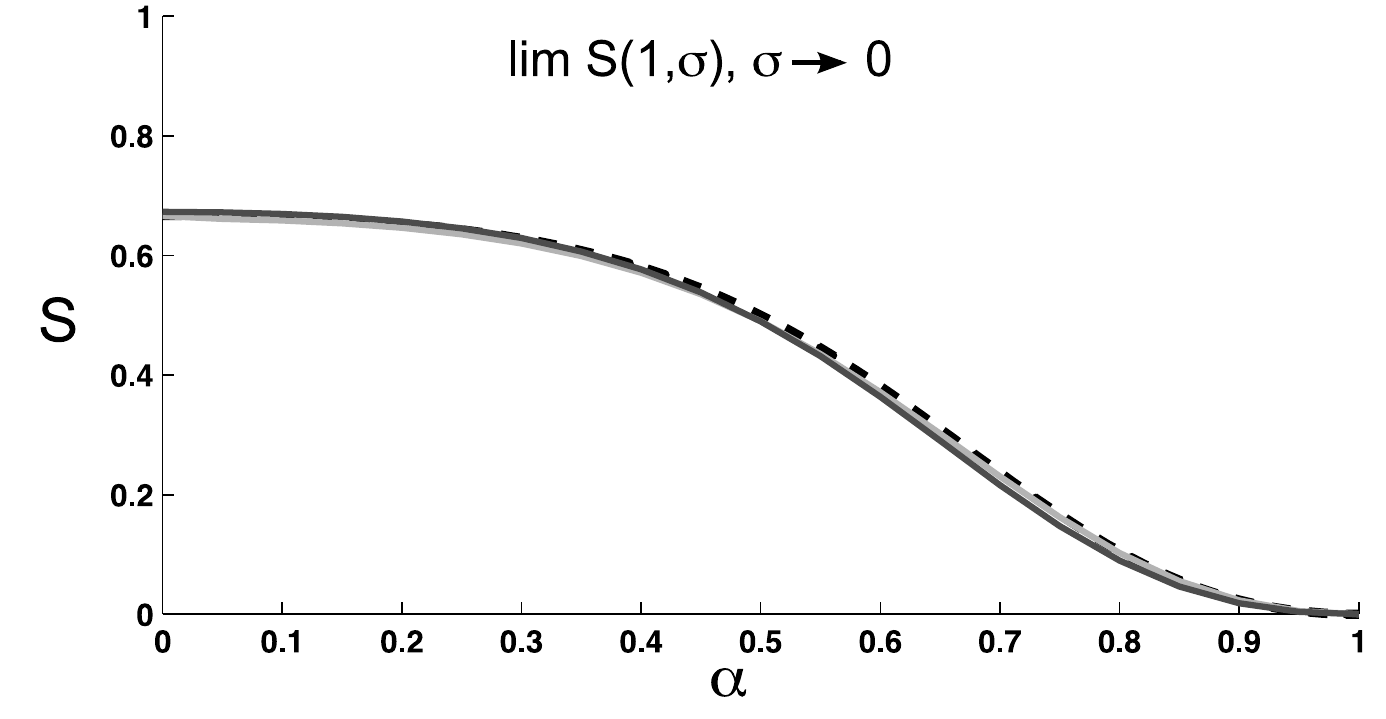}
\end{center}
\caption{$\lim_{\sigma \rightarrow 0^{+}} S(1,\sigma)$ for a continuum of models $\alpha \in [0,1]$ (black dashed). 
This is a good approximation for small ($\sigma = 0.2$; dark gray) and moderate ($\sigma = 1$; light gray) values of $\sigma$.}
\label{fig:sigma_approx_0}
\end{figure}
While this calculation is in the limit $\sigma \rightarrow 0^{+}$, it in fact remains a good approximation
for moderate $\sigma$, in fact, even for $\sigma \sim \omega$. Figure  \ref{fig:sigma_approx_0}
shows the limiting value as well as computed $S$ values at small (relative to $\omega = 1$; $\sigma = 0.2$) and moderate
($\sigma = 1$) values of $\sigma$. The limiting value remains a good approximation throughout this range.

\subsection*{An argument for general PRCs}

We close this by section by noting that, for arbitrary PRCs $z(\tht)$ and small $\sigma$, $$ d\nu/d\mu \propto \int_0^{2 \pi} Z(\tht)\, d\tht \; .$$
The calculations showing this are in the appendix.  While $S(\omega,\sigma) = \frac{\sigma^2 (d\nu/d\mu)^2}{CV^2 \nu}$ has other terms that depend on the PRC, this calculation does suggest that $S$ is likely to be smaller for PRCs with lower means in general, and lends some intuition to what drives the decrease in $S$ with $\alpha$.

\section{Correlation over shorter timescales}
 \begin{figure}[ht]
\begin{center}
\includegraphics[height=3in]{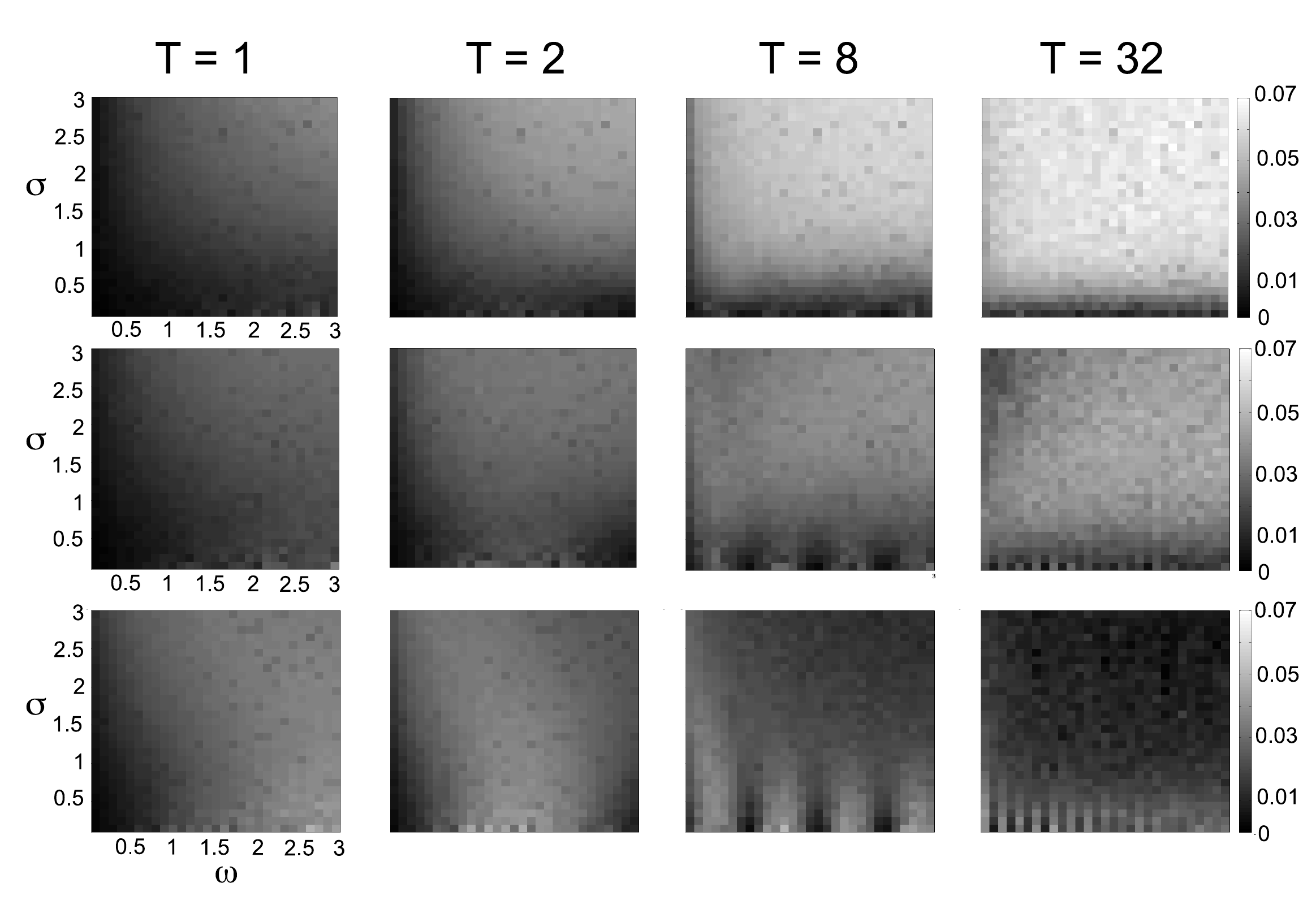}
\end{center}
\caption{$\rho_T(\omega,\sigma)$ as measured from Monte Carlo simulations for 
an increasing sequence of $T$, for simulations with $\alpha = 0$ (top row), $\alpha = 0.5$ (middle)
and $\alpha = 1$ (bottom). The fraction of common variance is $c = 0.1$; therefore, to recover the approximate
$S_T(\omega,\sigma) \approx \rho_T(\omega,\sigma)/c$, multiply by 10.  The $\omega$ and $\sigma$ axes of each plot are the same.
As $T$ increases, $S_T$ approaches the $T \rightarrow \infty$ limit illustrated in Fig. \ref{fig:rho_inflimit}.}
\label{fig:rho_T}
\end{figure}

 Figure \ref{fig:rho_T} shows $\rho_T$ computed for a sequence of finite time windows $T$. 
We can characterize the non-dimensional $T$ in terms of its length in terms
of a typical interspike interval (ISI) of the oscillator. The time window $T=1$ varies from
$0.05$ - $0.5$ ISI, roughly, from left to right; the time window $T=32$
varies from $1.6 - 16$ ISI. 

We see a striking
dependence of transferred correlations on $T$. $\rho_{32}$ is larger for Type I than for Type II oscillators
for most parameter values, consistent with our long time results. 

However, $\rho_1(\omega,\sigma)$ is smaller 
for Type I than for Type II for 95 \% of parameter pairs $(\omega,\sigma)$. This is consistent 
with recent results on the response of phase oscillators to correlated noise \cite{me08,GalanEU08}; 
in particular, \cite{me08} study the distribution of phase difference $\Delta \theta \equiv \theta_1 - \theta_2$ between two
oscillators driven by common noise and find that the probability that $\Delta \theta = 0$ is greater for Type II than for Type I.
As noted in the Introduction, this metric can be shown to have a direct relationship with our $\rho_T$ as $T \rightarrow 0$. 
To summarize, the ``switch" in $S_T(\omega,\sigma)$ (from higher correlation gain in Type II models to higher correlation gain in Type I) occurs at
time scales $T$ over which each cell fires several spikes; such timescales are biologically relevant, as we discuss in the next section.

\section{Summary and discussion}

We asked how correlated input currents are transferred into correlated spike trains in a class of nonlinear phase models that are generic reductions of neural oscillators.  Linear response methods, asymptotics, and Monte Carlo simulations gave the following answers:

\ben

\item  Over long timescales, Type I oscillators transfer ~66\% of incoming current correlations into correlated spike counts, while Type II oscillators transfer almost {\it none} of their input correlations into spike count correlations.  Models with intermediate phase response curves transfer intermediate levels of correlations.

\item Over long timescales, correlation transfer in Type I and Type II models is independent of the rate and coefficient of variation (CV) of spiking.  For intermediate models, correlation transfer decreases with CV and shows no clear dependence on rate.

\item That there is a timescale $T$ beneath which these results {\it reverse:} Type II neurons become more efficient at transferring correlations than Type I, there is an increasing dependence of correlation transfer on spike rate, and the strong dependence on CV weakens.

\een
We note that results (1) and (2) are highly distinct from findings for the leaky integrate-and-fire neuron model, for which up to 90-100\% of correlations are transferred over long timescales, with this level depending strongly on firing rate but very weakly on CV~\cite{rdsbjr07,brown07}.  This demonstrates a strong role for subthreshold nonlinearities in determining correlation transfer in the oscillatory regime (as seen for the quadratic integrate-and-fire model in~\cite{brown07}).

What timescales of spike count correlation actually matter in a given application?  This depends on the circuit that is ``downstream" of the pair (or, similarly, layer) of neural oscillators that we have studied in this paper; in other words, on what system is {\it reading out} the neurons we study here.  Clearly, different neurons and networks are sensitive to input fluctuations over widely varying timescales.  For example, some single neurons and circuits can respond only to events in which many of the cells that provide inputs spike nearly simultaneously.  This is the coincidence detector mode of operation (cf.~\cite{rd03} and references therin), and can result from fast membrane time constants (as occur in high-conductance states~\cite{DestexheRP03}); circuit mechanisms, such as feed-forward inhibition~\cite{PouilleS01}, can also play a role.  For such systems, short-timescale correlations among upstream cells are relevant -- small window lengths $T$.  On the opposite extreme, networks operating as neural integrators will accumulate inputs over arbitrarily long timescales (see \cite{brk03} and references therein); in this case, spike-time correlations over large windows $T$ are reflected in circuit activity.  In general there is a range of possible behaviors, and the timescales over which inputs are integrated can differ among various components of a network, among components of an individual cell~\cite{PouilleS01,rd03},
or among different times in a cell lifetime, depending on background input characteristics~\cite{DestexheRP03}.

One domain in which different levels of correlation transfer -- and different dependences of this correlation transfer on neurons' operating ranges -- can have a strong effect is the population coding of sensory stimuli.  For example, if neurons are read out over long timescales, then Type I vs Type II populations offer a choice between relatively high and low levels of correlation across the population.  Depending on heterogeneity of the population response to the stimulus at hand, one or the other of these choices can yield dramatically greater (Fisher) information about the encoded stimulus~\cite{zohary94,abbott99,sompolinsky01}.  The opposite choice of neuron type would be preferred for readout over short timescales, where trends in correlation transfer reverse.  Beyond averaged levels of correlation, a separate question that can affect encoding is whether correlations depend on the stimulus~\cite{PanzeriSTR99,josicSRD09}.  A natural way that this can occur is when correlations depend on the evoked rate or CV of firing.  We demonstrate that such dependencies are present in phase models over short timescales and, over long timescales, that they are present in intermediate but not ``purely" Type-I or Type-II models.  Once again, depending on details of stimulus encoding, these dependencies can either enhance or degrade encoding.  Overall, the picture that emerges is that correlation transfer is another factor to consider in asking which nonlinearities allow neuron models to best encode stimuli, and that there will be different answers for different stimuli.

In closing, we note that we have studied only simple (but widely-used) one-dimensional neural models here.  However, preliminary simulations suggest that the trends for correlation transfer found here also hold in some standard Type-I vs. Type-II conductance-based neuron models~\cite{excit}.  The situation is more complex, as global features of the neural dynamics can be involved, and will be explored in future work.


\newpage

\section{Appendix}
In this appendix we give some more details about calculation of $T_1(0)$, $T_2(0)$, and $\frac{dT_1}{d\mu}(0)$ for specific
values of $\alpha$.

\subsection{$\alpha = 1$ and numerical details}
For $Z(x) = -\sin(x)$ ($\alpha = 1$),
\begin{eqnarray}
A(x) &= & \omega + \frac{\sigma^2}{2} \sin(x) \cos(x)\\
B(x) &=& \sigma^2 [\sin(x)]^2
\end{eqnarray}
are periodic on $[0,\pi]$ with $B(x) = 0$ at $0 (2\pi)$ and $\pi$. 

$\Psi(x)$ is defined as an anti-derivative as in Eqn. (\ref{eqn:Psi}); we can 
compute this symbolically yielding
\begin{equation}
\Psi(x) = \sin(x) \exp(-\frac{2\omega}{\sigma^2} \cot(x)).
\end{equation}
We can check that $\lim_{x \rightarrow 0^+} \Psi(x) = 0$, $\lim_{x \rightarrow \pi^-} \Psi(x) = \infty$, and that $\Psi/B$ is not integrable at either $0$ or $\pi$.\\
For the interval $\pi$ to $2\pi$ we use the fact that $A(x)$ and $B(x)$ are $\pi$-periodic; $\Psi$ repeats on this second interval; that is
\begin{equation}
\Psi(x) = |\sin(x)| \exp \left( -\frac{2\omega}{\sigma^2} \cot(x) \right)
\end{equation} 
is an expression that is valid everywhere $Z(\theta) \not= 0$.

To compute values of $\frac{\partial T_1}{\partial x}$ on a uniform mesh, we use an adaptive quadrature routine
to evaluate Eqns. (\ref{eqn:dTidx1});
in particular Simpson's rule implemented via the MATLAB routine \textbf{quad}. We have already
demonstrated the integrability of Eqns. (\ref{eqn:dTidx1}) in \S \ref{sec:moments}.
$T_1(0)$ is then computed using Simpson's 3-point rule.

To compute values of $\frac{\partial T_2}{\partial x}$, we use adaptive quadrature as well, with the caveat
that $\frac{\partial T_1}{\partial x}$ is evaluated in between mesh points by linear interpolation.

\subsection{$\alpha = 0$}

For the theta model, $Z(x) = 1-\cos(x)$ or $\alpha = 0$,
\begin{eqnarray}
A(x) &=& \omega + \frac{\sigma^2}{2} (1-\cos(x)) \sin(x)\\
B(x) &=& \sigma^2 [1-\cos(x)]^2
\end{eqnarray}
are periodic on $[0,2\pi]$ with $B(x) = 0$ at $0 (2\pi)$.
Again we can integrate $2A(x)/B(x)$ symbolically, finding
\begin{equation}
\Psi = |1-\cos(x)| \exp \left( -\frac{2\omega}{3 \sigma^2} \frac{(2-\cos(x))\sin(x)}{(1-\cos(x))^2} \right)
\end{equation}

We can verify Eqn. (\ref{eqn:lHop_cond}) as before.

\subsubsection{A small $\sigma$ argument for general PRC}
Here we replicate the small $\sigma$ value of $d\nu/d\mu$, using a
perturbation expansion valid for an arbitrary PRC.

We consider the stationary densities $p(\theta)$ and $\hat{p}(\theta)$ of two different processes,
\begin{eqnarray}
d\theta & = & a(\theta) dt + b(\theta) dW_t\\
d\hat{\theta} & = & \hat{a}(\theta) dt + b(\theta) dW_t
\end{eqnarray}
where 
\begin{eqnarray}
a(\theta) & = & \omega + \frac{\sigma^2}{2} Z(\theta) Z'(\theta)\\
\hat{a}(\theta) & = & \omega + \frac{\sigma^2}{2} Z(\theta) Z'(\theta) + \mu Z(\theta)
\end{eqnarray}
$p(\theta)$ satisfies the stationary Fokker-Planck equation
\begin{eqnarray}
-\frac{\partial}{\partial \theta}(ap) + \frac{\partial^2}{\partial \theta^2}(bp) & = & 0
\end{eqnarray}
This can be integrated and the constant of integration is equal to the firing rate:
\begin{eqnarray}
ap - \frac{\partial}{\partial \theta} (bp) & = & \nu
\end{eqnarray}
and therefore
\begin{eqnarray} 
\nu & = & \frac{1}{2\pi}\int_0^{2\pi} a(\theta) p(\theta) d\theta\\
\hat{\nu}& = &  \frac{1}{2\pi}\int_0^{2\pi} \hat{a}(\theta) \hat{p}(\theta) d\theta
\end{eqnarray}
The DC input response, $d\nu/d\mu$, can be computed by differentiating $\hat{\nu}$ with respect to $\mu$ 
and evaluating at $\mu = 0$. Therefore we expand $\hat{\nu}$ as a series in $\mu$ and take the first-order term.
\begin{eqnarray}
\hat{p} & = & p + \mu p_1 + O(\mu^2)\\
\hat{\nu} & = & \nu + \mu \nu_1 + O(\mu^2)
\end{eqnarray}
We find that
\begin{eqnarray}
\frac{d\nu}{d\mu} = \nu_1 & = & \frac{1}{2\pi} \left[ \int_0^{2\pi} p(\theta) Z(\theta) d\theta +  \int_0^{2\pi} p_1(\theta) a(\theta) d\theta \right]\label{eqn:small_mu}
\end{eqnarray}

We consider (\ref{eqn:small_mu}) more carefully. We will see that all but one term are multiplied by $\sigma^2$; for small $\sigma$,
the term that remains significant is the mean of the phase-resetting curve.

We rewrite Eqn. (\ref{eqn:small_mu}) as follows:
\begin{eqnarray}
\nu_1 & = & \frac{1}{2\pi} \left[ \int_0^{2\pi} \left( \frac{1}{2\pi} + \tilde{p} (\theta) \right) Z(\theta) d\theta +  \sigma^2 \int_0^{2\pi} \frac{1}{2} Z(\theta)Z'(\theta) p_1(\theta) d\theta \right] \label{eqn:small_sigma}\\
 & = & \frac{1}{2\pi} \left[ \int_0^{2\pi} \frac{1}{2\pi} Z(\theta) d\theta  + \int_0^{2\pi} \tilde{p} (\theta)  Z(\theta) d\theta +  \sigma^2 \int_0^{2\pi} \frac{1}{2} Z(\theta)Z'(\theta) p_1(\theta) d\theta \right]  \label{eqn:small_sigma_2}
 \end{eqnarray}
 using the fact that $p_1(\theta)$ must average to $0$ so that $\hat{p}$ remains a probability density. We have also written $p(\theta)$, the stationary density for the process with $\mu = 0$, as the sum of a uniform density and a deviation $\tilde{p}(\theta)$. 
 
 Consider the process
 \begin{eqnarray}
 d\theta & = & \left( \omega + \frac{\sigma^2}{2} Z(\theta)Z'(\theta) \right) dt + \sigma Z(\theta) dW_t
 \end{eqnarray}
 The stationary density $p$ and firing rate $J$ satisfy
 \begin{eqnarray}
 \left( \omega + \frac{\sigma^2}{2} Z(\theta)Z'(\theta) \right) p(\theta) - \frac{\partial}{\partial \theta} \left( \frac{\sigma^2}{2} Z^2(\theta)p(\theta) \right) & = & J
 \end{eqnarray}
 If $\sigma = 0$ then the stationary density is $p_0 = \frac{1}{2\pi}$ and $J_0 = \frac{\omega}{2\pi}$.
 We expand $p$ and $J$ in powers of $\sigma^2$;
 \begin{eqnarray}
 p(\theta) & = & p_0 + \tilde{p}\\
 & = & p_0 + \sigma^2 \tilde{p}_1 + O(\sigma^4)\\
 J & = & J_0 + \sigma^2 J_1 + O(\sigma^4)
 \end{eqnarray}
 At $O(1)$ we have
 \begin{eqnarray}
 p_0 & = & \frac{J_0}{\omega}
 \end{eqnarray}
 as already stated. At $O(\sigma^2)$
 \begin{eqnarray}
 \tilde{p}_1 & = & \frac{1}{\omega} \left( J_1 + \frac{Z^2}{2} \frac{\partial p_0}{\partial \theta} + \frac{1}{2}Z Z' p_0 \right)\\
 & = & \frac{1}{\omega} \left( J_1 + \frac{1}{4\pi} Z Z' d \theta \right)
 \end{eqnarray}
 $J_1$ is determined so that $p$ is a probability density at any order:
 \begin{eqnarray}
 \int_0^{2\pi} J_1 = 2 \pi J_1 & = & -\frac{1}{4\pi} \int_0^{2\pi} Z Z' d\theta = 0
 \end{eqnarray}
 as $Z Z'$ is the perfect derivative of a periodic function. 
 For general order, we have
 \begin{eqnarray}
 \tilde{p}_n & = & \frac{1}{\omega} \left( J_n - \frac{1}{2} Z Z'  \tilde{p}_{n-1} + \frac{\partial}{\partial \theta} \left( \frac{Z^2}{2} \tilde{p}_{n-1} \right) \right)\\
 J_n & = & \frac{1}{4\pi} \int_0^{2\pi} Z Z' \tilde{p}_{n-1} d\theta
 \end{eqnarray}
 where we no longer expect $J_n$ to be zero.
 
 Let's return to (\ref{eqn:small_sigma_2}). The first integral is the mean of $Z(\theta)$. The second integral, in our expansion, first appears
 at fourth-order in $\sigma$. To see this, we examine
 \begin{eqnarray}
 \int_0^{2\pi} \tilde{p}_1 Z(\theta) d\theta & = & \frac{1}{4 \pi} \int_0^{2\pi} Z^2(\theta) Z'(\theta) d\theta \\
 & = & 0
 \end{eqnarray}
 The third term is second-order in $\sigma$.  Thus the dominant term in (\ref{eqn:small_sigma_2}) is the first one, and we have shown the desired result.
 
\bibliographystyle{plain}

\begin{thebibliography}{10}

\bibitem{abbott99}
L.~F. Abbott and P.~Dayan.
\newblock {The effect of correlated variability on the accuracy of a population
  code}.
\newblock {\em Neural Comput}, 11(1):91--101, 1999.

\bibitem{ad99}
L.F. Abbott and P.~Dayan.
\newblock The effect of correlated variability on the accuracy of a population
  code.
\newblock {\em Neural Computation}, 11:91--101, 1999.

\bibitem{alp06}
B.B. Averback, P.E. Latham, and A.~Pouget.
\newblock Neural correlations, population coding and computation.
\newblock {\em Nature Reviews Neuroscience}, 7:358--366, 2006.

\bibitem{Bai+01}
Wyeth Bair, Ehud Zohary, and William~T. Newsome.
\newblock {Correlated Firing in Macaque Visual Area MT: Time Scales and
  Relationship to Behavior}.
\newblock {\em J. Neurosci.}, 21(5):1676--1697, 2001.

\bibitem{biederlack07}
J.~Biederlack, M.~Castelo-Branco, S.~Neuenschwander, D.~W. Wheeler, W.~Singer,
  and D.~Nikoli\'c.
\newblock {{B}rightness induction: rate enhancement and neuronal
  synchronization as complementary codes}.
\newblock {\em Neuron}, 52(6):1073--1083, 2006.

\bibitem{Bin+01}
M.~D. Binder and R.~K. Powers.
\newblock {Relationship Between Simulated Common Synaptic Input and Discharge
  Synchrony in Cat Spinal Motoneurons}.
\newblock {\em J Neurophysiol}, 86(5):2266--2275, 2001.

\bibitem{bleistein}
N.~Bleistein and R.A. Handelsman.
\newblock {\em Asymptotic expansions of integrals}.
\newblock Dover, 1986.

\bibitem{britten92}
K.~H. Britten, M.~N. Shadlen, W.~T. Newsome, and J.~A. Movshon.
\newblock {The analysis of visual motion: a comparison of neuronal and
  psychophysical performance}.
\newblock {\em J Neurosci}, 12(12):4745--4765, 1992.

\bibitem{BHMphase}
E.~Brown, J.~Moehlis, and P.~Holmes.
\newblock On the phase reduction and response dynamics of neural oscillator
  populations.
\newblock {\em Neural Comp.}, 16:673--715, 2004.

\bibitem{brk03}
Brody C.D., Romo R., and Kepecs A.
\newblock Basic mechanisms for graded persistent activity: discrete attractors,
  continuous attractors, and dynamic representations.
\newblock {\em Current Opinion in Neurobiology}, 13:204--211, 2003.

\bibitem{chacron08}
M.~J. Chacron and J.~Bastian.
\newblock {Population Coding by Electrosensory Neurons}.
\newblock {\em J Neurophys}, 99(4):1825--1835, 2008.

\bibitem{chen06}
Y.~Chen, W.~S. Geisler, and E.~Seidemann.
\newblock {{O}ptimal decoding of correlated neural population responses in the
  primate visual cortex}.
\newblock {\em Nat Neurosci}, 9(11):1412--1420, 2006.

\bibitem{Cox+66}
D.R. Cox and P.A.W. Lewis.
\newblock {\em The Statistical Analysis of a Series of Events}.
\newblock Jonh Wiley, London, 1966.

\bibitem{rdsbjr07}
J.~de~la Rocha, B.~Doiron, E.~Shea-Brown, K.~Josic, and A.~Reyes.
\newblock Correlation between neural spike trains increases with firing rate.
\newblock {\em Nature}, 448, 2007.

\bibitem{deC+96}
R.~C. deCharms and M.~M. Merzenich.
\newblock Primary cortical representation of sounds by the coordination of
  action potentials.
\newblock {\em Nature}, 381:610--613, 1996.

\bibitem{DestexheRP03}
Alain Destexhe, Michael Rudolph, and Denis Par{\'e}.
\newblock The high-conductance state of neocortical neurons in vivo.
\newblock {\em Nature Reviews Neuroscience}, 4:739--751, 2003.

\bibitem{EI}
G.B. Ermentrout.
\newblock Type {I} membranes, phase resetting curves, and synchrony.
\newblock {\em Neural Comp.}, 8:979--1001, 1996.

\bibitem{ermentrout84}
G.B. Ermentrout and N.~Kopell.
\newblock {Frequency Plateaus in a Chain of Weakly Coupled Oscillators, I.}
\newblock {\em SIAM Journal on Mathematical Analysis}, 15:215, 1984.

\bibitem{GalanEU08}
Roberto~F. Gal‡n, G.~Bard Ermentrout, and Nathaniel~N. Urban.
\newblock Stochastic dynamics of uncoupled neural oscillators: Fokker-planck
  studies with the Þnite element method.
\newblock {\em Phys. Rev. E}, 76:056110, 2007.

\bibitem{gardiner}
C.W. Gardiner.
\newblock {\em Handbook of Stochastic Methods}.
\newblock Series in Synergetics. Springer, 3rd edition, 2004.

\bibitem{Gra+89}
C.~M. Gray, P.~K\"{o}ing A.~K. Engel, and W.~Singer.
\newblock Oscillatory responses in cat visual cortex exhibit inter-columnar
  synchronization which reflects global stimulus properties.
\newblock {\em Nature}, 338:334--337, 1989.

\bibitem{HMM93}
D.~Hansel, G.~Mato, and C.~Meunier.
\newblock Phase dynamics for weakly coupled {H}odgkin-{H}uxley neurons.
\newblock {\em Europhys. Lett.}, 25(5):367--372, 1993.

\bibitem{johnson80}
K.~O. Johnson.
\newblock {Sensory discrimination: neural processes preceding discrimination
  decision}.
\newblock {\em J Neurophys}, 43(6):1793--1815, 1980.

\bibitem{josicSRD09}
K.~Josic, E.~Shea-Brown, B.~Doiron, and J.~de~la Rocha.
\newblock Stimulus-dependent correlations and population codes.
\newblock {\em Neural Computation}, 2009.
\newblock In Press.

\bibitem{kohn05}
A.~Kohn and M.~A. Smith.
\newblock {{S}timulus dependence of neuronal correlation in primary visual
  cortex of the macaque}.
\newblock {\em J Neurosci}, 25(14):3661--3673, 2005.

\bibitem{kohnprep}
A.~Kohn, M.~A. Smith, and J.~A. Movshon.
\newblock Effect of prolonged and rapid adaptation on correlation in {V}1.
\newblock {\em Computational and Systems Neuroscience, Cold Spring Harbor NY
  (abstract)}, 2004.

\bibitem{KuhnAR03}
A.~Kuhn, A.~Aertsen, and S.~Rotter.
\newblock Higher-order statistics of input ensembles and the response of simple
  model neurons.
\newblock {\em Neural Comp.}, 15:67--101, 2003.

\bibitem{TetzlaffRSAD08}
A.~Kuhn, A.~Aertsen, and S.~Rotter.
\newblock Dependence of neuronal correlations on filter characteristics and
  marginal spike train statistics.
\newblock {\em Neural Comp.}, 20:2133--2185, 2008.

\bibitem{ldl05}
B.~Lindner, B.~Doiron, and A.~Longtin.
\newblock Theory of oscillatory firing induced by spatially correlated noise
  and delayed inhibitory feedback.
\newblock {\em Physical Review E}, 72, 2005.

\bibitem{lindner03}
B.~Lindner, A.~Longtin, and A.~Bulsara.
\newblock Analytic expressions for rate and {CV} of a type {I} neuron driven by
  gaussian write noise.
\newblock {\em Neural Computation}, pages 1761--1780, 2003.

\bibitem{me08}
S.~Marella and G.B. Ermentrout.
\newblock Class-{II} neurons display a higher degree of stochastic
  synchronization than class-{I} neurons.
\newblock {\em Physical Review E}, 77, 2008.

\bibitem{Mor+06}
R.~Moreno-Bote and N.~Parga.
\newblock Auto- and crosscorrelograms for the spike response of leaky
  integrate-and-fire neurons with slow synapses.
\newblock {\em Phys. Rev. Lett.}, 96:028101, 2006.

\bibitem{oram98}
M.~W. Oram, P.~F{\"o}ldi{\'a}k, D.~I. Perrett, and F.~Sengpiel.
\newblock {The `Ideal Homunculus': decoding neural population signals}.
\newblock {\em Trends Neurosci}, 21(6):259--265, 1998.

\bibitem{PanzeriSTR99}
S.~Panzeri, S.~Schultz, A.~Treves, and E.~T. Rolls.
\newblock Correlations and the encoding of information in the nervous system.
\newblock {\em Proc Royal Soc Lond B}, 266:1001--1012, 1999.

\bibitem{pmgh05}
B.~Pfeuty, G.~Mato, D.~Golomb, and D.~Hansel.
\newblock The combined effects of inhibitory and electrical synapses in
  synchrony.
\newblock {\em Neural Computation}, 17:633--670, 2005.

\bibitem{roelfsema08}
J.~Poort and P.~R. Roelfsema.
\newblock Noise correlations have little influence on the coding of selective
  attention in area v1.
\newblock {\em Cerebal Cortex}, 2008.
\newblock Advanced Online Publication.

\bibitem{PouilleS01}
Frederic Pouille and Massimo Scanziani.
\newblock Enforcement of temporal fidelity in pyramidal cells by somatic
  feed-forward inhibition.
\newblock {\em Science}, 293:1159--1164, 2001.

\bibitem{excit}
J.~Rinzel and G.B. Ermentrout.
\newblock Analysis of neural excitability and oscillations.
\newblock In C.~Koch and I.~Segev, editors, {\em Methods in Neuronal Modeling},
  pages 251--291. MIT Press, 1998.

\bibitem{romo03}
R.~Romo, A.~Hernandez, A.~Zainos, and E.~Salinas.
\newblock {Correlated neuronal discharges that increase coding efficiency
  during perceptual discrimination}.
\newblock {\em Neuron}, 38(4):649--657, 2003.

\bibitem{rd03}
M.~Rudolph and A.~Destexhe.
\newblock Tuning neocortical pyramidal neurons between integrators and
  coincidence detectors.
\newblock {\em Journal of Computational Neuroscience}, 14:239--251, 2003.

\bibitem{ss00}
E.~Salinas and T.J. Sejnowski.
\newblock Impact of correlated synaptic input on output firing rate and
  variability in simple neuronal models.
\newblock {\em Journal of Neuroscience}, 20(16):6193--6209, 2000.

\bibitem{samonds03}
J.~M. Samonds, J.~D. Allison, H.~A. Brown, and A.~B. Bonds.
\newblock {Cooperation between Area 17 Neuron Pairs Enhances Fine
  Discrimination of Orientation}.
\newblock {\em J Neurosci}, 23(6):2416, 2003.

\bibitem{schneidman06}
E.~Schneidman, M.~J. Berry, R.~S. II, and W.~Bialek.
\newblock {Weak pairwise correlations imply strongly correlated network states
  in a neural population}.
\newblock {\em Nature}, 440(7087):1007, 2006.

\bibitem{SeriesLP04}
P.~Seri\`es, P.~E. Latham, and A.~Pouget.
\newblock Tuning curve sharpening for orientation selectivity: coding
  efficiency and the impact of correlations.
\newblock {\em Nat Neurosci}, 7:1129Ð1135, 2004.

\bibitem{sn98}
M.N. Shadlen and W.T. Newsome.
\newblock The variable discharge of cortical neurons: Implications for
  connectivity, computation, and information coding.
\newblock {\em Journal of Neuroscience}, 18(10):3870--3896, 1998.

\bibitem{shamir04}
M.~Shamir and H.~Sompolinsky.
\newblock {Nonlinear population codes}.
\newblock {\em Neural Comput}, 16(6):1105--1136, 2004.

\bibitem{shamir06}
M.~Shamir and H.~Sompolinsky.
\newblock {{I}mplications of neuronal diversity on population coding}.
\newblock {\em Neural Comput}, 18(8):1951--1986, 2006.

\bibitem{brown07}
E.~Shea-Brown, K.~Josi\'{c}, B.~Doiron, and J.~de~la Rocha.
\newblock Correlation and synchrony transfer in integrate-and-fire neurons:
  Basic properties and consequences for coding.
\newblock {\em Phys Rev Lett}, 100:108102, 2008.

\bibitem{sompolinsky01}
H.~Sompolinsky, H.~Yoon, K.~Kang, and M.~Shamir.
\newblock {Population coding in neuronal systems with correlated noise}.
\newblock {\em Phys Rev E}, 64(5 Pt 1):051904, 2001.

\bibitem{geomtime}
A.~Winfree.
\newblock {\em The Geometry of Biological Time}.
\newblock Springer, New York, 2001.

\bibitem{zohary94}
E.~Zohary, M.~Shadlen, and W.~Newsome.
\newblock Correlated neuronal discharge rate and its implications for
  psychophysical performance.
\newblock {\em Nature}, 370:140--143, 1994.

\end{thebibliography}


\end{document}